\documentclass{article}
\usepackage{graphicx}
\usepackage{amsmath, amssymb}
\usepackage{bm}
\usepackage{cases}
\usepackage{slashed}
\usepackage{physics}
\usepackage{enumitem}
\usepackage{hyperref}
\usepackage{dsfont}
\usepackage{titlesec}
\usepackage{dsfont}
\usepackage{graphicx, xcolor, import}
\usepackage{float}
\usepackage{caption}
\usepackage{transparent}
\usepackage{tensor}
\usepackage{tcolorbox}
\usepackage{simpler-wick}
\usepackage{multirow}
\usepackage{blkarray}
\usepackage{empheq}
\usepackage{ragged2e}
\usepackage{authblk}
\newcommand{\myS}{Schr\"{o}dinger}

\usepackage [english]{babel}
\usepackage [autostyle, english = american]{csquotes}
\MakeOuterQuote{"}
\usepackage[margin=1in]{geometry}

\title{\textbf{Spontaneous Unitarity Violation and Quantum State Reduction}}
\author{Dakshansh Chawda}
\affil{\small \textit{University of Amsterdam, 1098 XH Amsterdam, The Netherlands}}
\date{}
\begin{document}
\maketitle

\vspace{2in}
\begin{abstract}
    One of the central foundational problems in Quantum Mechanics is the inability of Schr\"{o}dinger's unitary time evolution to describe measurement of a quantum state. Recent developments \cite{SUV},\cite{PT},\cite{BTTS} in models of spontaneous unitarity violation (SUV) propose that quantum state reduction can emerge as a thermodynamic phenomenon, offering a natural resolution to the measurement problem. This paper investigates the philosophical implications of SUV in relation to the nature of probabilities in quantum mechanics and statistical mechanics. It provides an alternate interpretation of probabilities in quantum mechanics, that aligns with (and improves) the arguments of the propensity interpretation of probabilities. 
\end{abstract}

\newpage
\tableofcontents


\newpage
\section{Introduction}
Quantum mechanics is a well-established theory that agrees with experimental tests. It explains phenomena from the scale of individual atoms ($10^{-10}$  m length scale) to larger systems involving aggregates of atoms. While quantum mechanics (QM) is foundational at this scale, macroscopic behavior is often described by emergent theories like thermodynamics or classical mechanics derived from quantum mechanical principles. At the mesoscopic scale, the quantum and classical effects overlap, leading to phenomena like decoherence. The mesoscopic regime encounters a conflict between the well-established theories of quantum and classical mechanics when the macroscopic measuring device measures a microscopic quantum system. This conflict is one of foundational problems of quantum mechanics - The Measurement Problem.\\

Following sections describe the measurement problem and proposals for solving it. One such proposal is objective collapse theories (OCTs), which lead to dynamical quantum state reduction (DQSR) models, which is discussed further.\par

Sec. 2 and 3 aim to provide a brief introduction to the essential features of SUV. It draws a lot from \cite{SUV}-\cite{Lenstra}, and only an overview of calculations and the essential results are provided, skipping hefty derivations; for which the reader can refer to the existing literature on the subject. Sec. 2 provides the theoretical background on standard QM framework, spontaneous symmetry breaking (SSB) and the breaking of time translation symmetry. Sec. 3 delves into Spontaneous Unitarity Violation (SUV), followed by Sec. 4 and Sec. 5 that discuss the philosophical foundations and the philosophical implications of SUV, specifically the interpretation of the nature of probabilities in QM. 
Finally, Sec. 6 concludes with the key philosophical insights provided by SUV models and how it changes QM, as we know it. 

\subsection{The Measurement Problem}
The measurement problem in quantum mechanics is the apparent conflict between the deterministic evolution of wave function (according to \myS's equation) and the probabilistic nature of measurement outcome (which obey Born's probability distribution).

To better formulate this, one can split the measurement process into two stages: Let the initial state of the combined apparatus-quantum system be 
\begin{equation}
|\Psi_{\text{initial}}\rangle = \left( \sum_i c_i |s_i\rangle \right) \otimes |A_0\rangle
\end{equation}

where, the apparatus is in the neutral state $\ket{A_0}$. The first stage is the "pre-measurement" part, where the quantum system becomes entangled with the apparatus, correlating the system's states $\ket{s_i}$ with apparatus states $\ket{A_i}$, called "pointer states" (in classical terms, it's analogous to the needle of a gauge pointing to a value): 
\begin{equation}
|\Psi_{\text{entangled}}\rangle = \sum_i c_i |s_i\rangle \otimes |A_i\rangle
\end{equation}
The state of the system and apparatus together is still in a superposition, reflecting quantum coherence \cite{Zurek}. Second stage is the "reduction" part, where this superposition reduces to just a single pointer state.\newline 
Due to interactions with the environment (e.g., photons or thermal noise), the apparatus states $\ket{A_i}$ lose coherence with one another. The environment entangles with the apparatus, and the total state becomes:

\begin{equation}
|\Psi_{\text{decohered}}\rangle = \sum_i c_i |s_i\rangle \otimes |A_i\rangle \otimes |E_i\rangle
\end{equation}
Upon measurement, the system is observed in one state $\ket{s_m}$ with the apparatus pointer indicating the corresponding outcome:
\begin{equation}
|\Psi_{\text{collapsed}}\rangle = |s_m\rangle \otimes |A_m\rangle
\end{equation}
So formulated, the measurement problem is about the inability of unitary time evolution prescribed by \myS's equation to explain the transition: 
\begin{equation}
\sum_i c_i\ket{s_i} \rightarrow \ket{s_m}
\end{equation} 

To address the measurement problem, the theoretical efforts made so far can be grouped into the following three categories:
\begin{enumerate}
    \item Decoherence: Well-grounded in the mathematical formalism of quantum mechanics \cite{Zurek},\cite{Zeh} and experimentally supported, decoherence explains the appearance of classical outcomes and eliminates the need for a unique "preferred basis". The idea is that the quantum state of a system gets entangled with the many degrees of freedom of the environment. The resulting density matrix is a pure state but too large to compute. So, one has to trace out the environmental degrees of freedom which cannot be measured. This causes the state to behave effectively classically. However, while decoherence explains how classical behavior arises from quantum systems, it does not resolve the measurement problem (why one specific outcome is observed). This is due to the fact that taking a partial trace over the environment degrees of freedom is equivalent to taking the expectation value of operators describing the unobserved state, which is defined within an ensemble, thereby, explicitly restricting one to describing ensemble averages. Thus, it complements interpretations rather than serving as an interpretation itself. It cannot resolve the measurement problem for a single outcome in a single measurement.
    
    \item Interpretations: This is the class of all interpretations of quantum mechanics, which assume the \myS's equation is correct and applies at all scale to all the objects. The most well-known of these is the Copenhagen interpretation.\\ Since all of these assume the universality of \myS's equation, they make identical empirical predictions and cannot be experimentally distinguished or verified. Also, these interpretations often replace the problem of measurement with some other (new) problem. For instance, in the Many-Worlds Interpretation eliminates wave function collapse but introduces the "preferred basis problem" - what determines the basis in which the branching of worlds occurs, and how do we recover classical reality from the multiverse \cite{MWI}? 

    \item Objective Collapse Theories: These theories assume the ontological nature of quantum state, and propose modifications to \myS's equations. These modifications have no noticeable effect at the microscopic level (so they are in agreement with the physics of elementary particles), but they influence the dynamics at the mesoscopic level. This accounts for the state reduction even for a single outcome in a single measurement, unlike decoherence. \par
    Since OCTs introduce refinements to \myS's equation, they can be tested experimentally. In the next section, we discuss the requirement for any OCT, also known as DQSR models.

\end{enumerate}

\subsection{Dynamical Quantum State Reduction}
In general, all OCTs lead to DQSR models with the following correction to \myS's equation: 
\begin{equation}
i\hbar \frac{\partial}{\partial t}\ket{\psi(t)} = (\hat{H}_0 + \epsilon \hat{\mathcal{G}})\ket{\psi(t)} \label{dqsr}
\end{equation} 
We now discuss the constraints on such models so that they adhere to the requirements of an OCT.\par
The first constraint is that any DQSR described as Eq.(6) should have a time evolution that reduces to standard quantum mechanics when dealing with systems in the microscopic regime. For this to hold, $\epsilon$ should be small enough to have no significant change at the microscopic level. However, the DQSR should also predict effectively instantaneous collapse of superposition of mesoscopic objects. This requires $\hat{\mathcal{G}}$ to scale with the some extensive quantity of the system $N$ so that the collapse time scale $\tau_c \propto 1/(N\epsilon)$ \cite{QD}.\\
The second constraint is that it should cause stable quantum state reduction, i.e., the superposition state of the system should reduce to a single pointer state and once reduced it should not spontaneously evolve out of it. This requires $G$ to be non-Hermitian and coupled to an order parameter of any symmetry-breaking system \cite{BTTS}. Finally, any such model should reproduce Born's rule. This necessitates a non-linear and a stochastic component \cite{BR}.\\

Therefore, the requirements can be summarised as: nonunitarity, nonlinearity, stochasticity, dependence on the order parameter of symmetry breaking system and suitably small strength $\epsilon$. 

\section{Theoretical Background}
This section includes a brief overview of the standard QM framework, SSB mechanism and its relation to singular limits, followed by a discussion of the breaking of time translation symmetry in QM.

\subsection{Standard Quantum Mechanics Framework}
For the sake of completeness, a brief review of standard QM framework is provided here. We quickly go over the unitary evolution and Born's rule, starting with the postulates of QM.

\subsubsection*{Postulates of Quantum Mechanics}
\begin{itemize}
    \item Postulate 1: State Description: In the standard framework of QM, any state of a closed quantum system is describe by a state vector $\ket{\psi}$ in a Hilbert space.
    \item Postulate 2: Evolution of States: The time evolution of a quantum state is governed by \myS's Equation: 
    \begin{equation}
        i\hbar\frac{\partial}{\partial t}\ket{\psi(t)} = \hat{H}\ket{\psi(t)},
    \end{equation}
    where $\hat{H}$ is the Hamiltonian operator of the system.
    \item Postulate 3: Measurement: Measurement outcomes correspond to the eigenvalues of the operator representing the observable. These observables are the physical quantities (e.g., position, momentum, energy) and are represented by Hermitian operators. After measurement, the system collapses into the eigenstate corresponding to the observed eigenvalue.
    \item Postulate 4: Born's Rule: The probability of obtaining a measurement outcome $o_m$ is given by:
    \begin{equation}
        P(o_m) = |\langle o_m|\ket{\psi}|^2,
    \end{equation}
    where $\ket{o_m}$ is the eigenstate associated with $o_m$.
\end{itemize}

\subsubsection*{Unitary Evolution}
\begin{itemize}
    \item The evolution of a quantum state is deterministic and governed by the unitary operator:
\begin{equation}
    \ket{\psi(t)} = \hat{U}(t)\ket{\psi(0)}, \quad \hat{U}(t) = \text{exp}\left(-\frac{i}{\hbar}\hat{H}t\right)
\end{equation}

    \item Unitarity ensures the conservation of probability at all times: $\bra{\psi(0)}\psi(0)\rangle = 1 = \bra{\psi(t)}\psi(t)\rangle$
\end{itemize}

\subsection{Spontaneous Symmetry Breaking and Singular limits}
Spontaneous Symmetry Breaking is the phenomenon in which a ground state solution of the equation of motion of a Lagrangian is not symmetric under a symmetry of the Lagrangian. The fact that this symmetry breaking is unpredictable and unavoidable in practice makes it `spontaneous'. It is important to note that in strictly finite-dimensional quantum systems, true SSB cannot occur due to the Stone-von Neumann theorem. This theorem implies that the representation of the canonical commutation relations is unique in finite dimensions, and the ground state of the system cannot exhibit distinct, degenerate symmetry-broken phases. However, finite systems (e.g., ferromagnetic materials) can exhibit approximate symmetry breaking as an emergent phenomenon that  becomes exact in the thermodynamic limit $N \rightarrow \infty$, where $N$ is the degrees of freedom \cite{QD}. In such a case, the order parameter defines the extent of approximate symmetry breaking.\\

The order parameter operator $\hat {\mathcal{O}}$, in general, does not commute with the Hamiltonian. So, in the case of a harmonic crystal the order parameter can be the position operator $\hat{X}$ which does not commute with the Hamiltonian: 
\begin{equation}
    \hat{H} = \frac{\hat{P}_{tot}^2}{2m_{tot}} + \epsilon N(\hat{X}_{com} - x_0)^2
\end{equation}
Here, $\hat{X}_{com}$ is the center of mass position operator, and $x_0$ is the centre of the externally applied potential, and $m_{tot} = mN$. Notice that the applied potential is couple to the order parameter $N$ of the crystal. So, in case there is an infinitesimally small perturbation, the ground state of the system, $\ket\psi_{gs}$, is a fully localised symmetry-broken state in the thermodynamic limit. However, if there is no external applied potential, $\ket\psi_{gs}$ is fully delocalised and symmetric. The spontaneous breakdown of translation symmetry can be understood as a consequence of non-commutative (or singular) nature of the limits:
\begin{align}
    \lim_{N\to\infty}\lim_{\epsilon\to 0} \ket\psi_{gs} &= \ket{P_{tot} = 0}\\
    \lim_{\epsilon\to 0}\lim_{N\to\infty} \ket\psi_{gs} &= \ket{X_{com} = x_0}
\end{align}
Though never realised for a finite size object, these singular limits represent a \textit{diverging susceptibility} of the crystal to symmetry-breaking perturbations \cite{iSSB}. For a large symmetric system a perturbation with an energy scale as small as $\sim 1/N$ is sufficient to force it into a symmetry-broken state. Therefore, for all practical puposes, there will always be such a perturbation that causes SSB in a harmonic crystal.
 
\subsection{Breaking Time Translation Symmetry}
Now that we have discussed the basic notion of SSB and singular limits, we discuss the breaking of time translation symmetry in QM. Just as homogeneity of space leads to translational invariance of Hamiltonian, homogeneity of time imposes time-translation symmetry. Under the application of an external potential, one can spontaneously break this translational invariance forcing the system into a symmetry-broken state. To break time translation symmetry we introduce a small unitarity breaking field into the generator of time evolution operator:
\begin{equation}
    \hat{U}(t) = \text{exp}\left(-\frac{i}{\hbar}t[\hat{H}_0-i\hat{\mathcal{O}}]\right)
\end{equation}
Here, $\hat{H}_0$ is the symmetric Hamiltonian of the unperturbed system. One might notice that there is a possible lack of energy conservation due to the non-unitary dynamics. But this is only a pseudo-problem as the non-Hermitian field only causes transitions between states which are degenerate in the thermodynamic limit, thus leaving the total energy untouched. Mukherjee and Wezel \cite{Aritro} show that this holds as long as the non-Hermitian field is coupled only to the order parameter and no internal degrees of freedom of the system.  

\section{Spontaneous Unitarity Violation: Models and Mechanisms}
In the previous sections, we discussed the framework of QM, SSB, broken time translation symmetry and some of its implications for QM formalism. In this section, we combine all these ingredients to form the general structure of SUV models. Later, we discuss the emergence of Born's rule in the framework of SUV models. 

\subsection{SUV: The Central Idea}
We already discussed some of the essential feature of a DQSR model in Sec. 1.2. Here, we make these requirements more explicit and form SUV model based these requirements.\\
Following Eq.\eqref{dqsr} and the requirement of nonunitarity, nonlinearity, stochastiticy and extensivity lead to the modified \myS's equation:
\begin{equation}
    i\hbar\frac{\partial}{\partial t}\ket{\psi(t)} = (\hat{H}_0 + i\epsilon\mathcal{N}\hat{\mathcal{G}})\ket{\psi(t)} \label{suv}
\end{equation}
where, $\hat{\mathcal{G}}$ is a hermitian operator, and $\mathcal{N}$ is the order parameter which is generally extensive. The origin of $\mathcal{N}$ in the above equation is discussed in \cite{SSB}. This factor of $\mathcal{N}$ ensures that for large systems, i.e., for large $\mathcal{N}$, the perturbation dominates the dynamics while for small systems, the free Hamiltonian $\hat{H}_0$ dominates the equation and thus the predictions are effectively indistinguishable from the original \myS's equation in standard QM.\\

Besides breaking unitarity we need to break a symmetry of the Hamiltonian. The wave function will then collapse to one of these symmetry-broken states. As a consequence, SUV does not encounter any "preferred basis problem" as mentioned in Sec 1.1. Since we discussed it in the context of the Many-Worlds Interpretation, it is worthwhile to state it in general, once - The preferred basis problem is about the supposed emergence of certain "preferred bases" when a measurement is made, like the position basis, momentum basis and spin-up/spin-down basis to name a few. One may ask why these bases are observed as the "preferred" or classical outcomes instead of arbitrary quantum superpositions. This issue is tied to the emergence of classicality. Often, this is explained in decoherence terminology as the result of the role the environment plays in selecting a stable basis. We briefly discussed this in Sec 1.1 when we discussed decoherence.\par 

Unlike the various interpretations of QM which often insert these bases artificially, here, due to SSB of a symmetry of the Hamiltonian we get these allowed bases by the symmetry-broken ground states of the perturbed system. In the next section we discuss the mechanism for this - Dynamical Spontaneous Symmetry Breaking.

\subsection{Dynamical Spontaneous Symmetry Breaking}
The Hamiltonian operator in quantum mechanics has a dual role:
\begin{enumerate}
    \item Spectral role: This refers to the Hamiltonian's function in determining the eigenstates and eigenvalues of the system, which correspond to the possible energy states the isolated quantum system can occupy.
    \item Dynamical role: This refers to the Hamiltonian's function as the generator of time evolution for a general quantum state, dictating how the state evolves over time.
\end{enumerate}
While the Equilibrium theory of SSB deals with the Spectral role in the thermodynamic limit, Dyanmical SSB addresses the Dynamical role in that limit - unitary time translation symmetry can be spontaneously broken down in the thermodynamic limit.\\

To begin with, we introduce a small, non-unitary perturbation to $\hat{H}_0$. For the harmonic crystal, this will lead to \cite{SUV}:
\begin{equation}
    \hat{H}_{SUV} = \frac{\hat{P}_{tot}^2}{2m_{tot}} + i\epsilon\mathcal{N}(\hat{X}_{com} - x_0)^2 \label{Hsuv}
\end{equation}
Now, just as in the case of SSB where we get a singular limit corresponding to Equilibrium SSB, here we find a singular limit for the Dynamic SSB of a given initial state:
\begin{align}
    \lim_{\mathcal{N}\to\infty}\lim_{\epsilon\to 0} e^{-\frac{i}{\hbar}\hat{H}_{SUV}t}\ket{P_{tot}=0} &= \ket{P_{tot} = 0} \label{l1} \\
    \lim_{\epsilon\to 0}\lim_{\mathcal{N}\to\infty} e^{-\frac{i}{\hbar}\hat{H}_{SUV}t}\ \ket{P_{tot}=0} &= \ket{X_{com} = x_0} \quad \forall t>0 \label{l2}
\end{align}
The above equations show that in the absence of any non-unitary perturbations, the quantum state evolves according to \myS's equation of Standard QM. However, even a vanishingly small non-unitary perturbation is enough to cause localisation of the wave function to the localisation centre $x_0$.\\

Again, the existence of the non-commuting, singular limit represent a \textit{diverging susceptibility} of the crystal to non-unitary, symmetry-breaking perturbations . For large crystals (say, $\mathcal{N} \sim 10^{23}$ atoms), the perturbation required to cause the `collapse' is of the order $1/\mathcal{N}$. This is so small so as to render it beyond the control of experimenter or any physical situation making it impossible for large, harmonic crystals (or any large system, in general) to avoid localisation. Since this localisation is unavoidable and the localisation centre $x_0$ unpredictable, in practice, the unitary time evolution can be said to be violated spontaneously.
\newpage

\subsection{Emergence of Born's rule}
We notice from the previous section that the spontaneous breakdown of unitarity "emerges" as the thermodynamic limit is approached. So, microscopic systems will not be spontaneously evolved out of a delocalised state. In fact, for a vanishingly small non-unitary perturbation to do so it will require longer that the age of the universe to have a measurable effect on the dynamics of the system \cite{SUV}.\\

For Born's rule to emerge from non-unitary dynamics one needs $\hat{\mathcal{G}}$ (in Eq. \eqref{suv}) to be nonlinear and stochastic. One such possibility is: 
\begin{equation}
\hat{\mathcal{G}}\ket{\psi(t)} = \biggl (J\langle \hat{\sigma}_z \rangle + G\xi(t) \biggr ) \biggl [\hat{\sigma}_z - \langle \hat{\sigma}_z \rangle \biggr ] \ket{\psi(t)} \label{pert}
\end{equation}
where $J$ is the self coupling constant of the nonlinear part of the system with units [$J$] = $s^{-1}$, and $G$ is the coupling constant of the stochastic field with the same units as $J$.
Mertens et al. \cite{BR} showed that if one requires Born's rule to emerge from a theory instead of being imposed then the theory cannot be linear. Whereas, the origin of stochastic component is essential for the collapse to be probabilistic. The stochastic field provides a way to introduce randomness and probabilistic effects into an otherwise deterministic model. Mertens et al. \cite{BR} worked out a two-state model based on Lieb-Mattis antiferromagnet (AFM) and defined a minimal example for the collapse dynamics as 
\begin{equation}
    d\theta_\xi (t) = -dt J\sin{\theta_\xi(t)}(\cos{\theta_\xi (t)} - \frac{G}{J}\xi(t))  \label{afm}
\end{equation}
where $\theta_\xi (t)$ is the polar angle corresponding to a general two-state system parameterized on the Bloch sphere. This equation describes possible dynamics of a collapsing state. One can visualise the evolution of the wave function on the Bloch sphere by flow lines. Under unitary time evolution, i.e., using standard QM time evolution operator $e^{-i\hat{H}t/\hbar}$, the flow on the Bloch sphere is conservative and consists of closed cycles known as Rabi oscillations. As long as the generator of time evolution is unitary the flow lines are closed loops and the flow is conservative, i.e., each initial state will undergo indefinite periodic time evolution as shown in figure 1.

\begin{figure}[htb]
    \centering
    \begin{minipage}{0.45\textwidth}
        \centering
        \includegraphics[width=\linewidth]{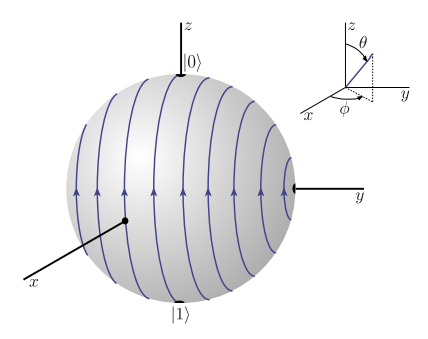}
        \captionsetup{width=\linewidth} 
        \caption{Unitary evolution on the Bloch sphere. The flow lines indicate Rabi oscillations \cite{BR}.}
    \end{minipage}%
    \hspace{0.5cm} 
    \begin{minipage}{0.40\textwidth}
        \centering
        \includegraphics[width=\linewidth]{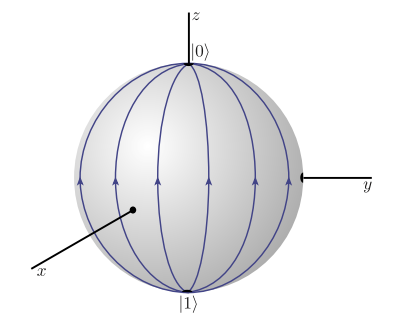}
        \captionsetup{width=\linewidth} 
        \caption{Flow generated by a
purely diagonal non-Hermitian generator of time evolution. The state $\ket{0}$ is an attractive fixed point of the flow, while $\ket{1}$
is a repulsive fixed point \cite{BR}.}
    \end{minipage}
\end{figure}

For an OCT, on the other hand, the system should end up in either one of the pointer states, and evolve no further then as depicted in figure 2. On the Bloch sphere this translates to the state ending up at either the north pole of the south pole in the Lieb-Mattis AFM example. The pole towards which the state evolves needs to be a stable end-point, i.e., either attractive fixed point or limit cycles. 
Now consider the two extreme cases:
\begin{enumerate}
    \item Ratio $G/J << 1:$ Then the nonlinear part of the system dominates and the state evolves to the nearest fixed point and never to the other, thus not reproducing Born's rule.
    \item Ratio $G/J >> 1:$ In this case the stochastic part takes over the evolution dynamics and every state would randomly evolve to one of the fixed points with equal probability (1/2 for each), for all initial states. Again, this does not reproduce Born's rule. 
\end{enumerate}
This implies that there is a specific value for this ratio of relative coupling strength $G/J$ that, as it turns out, depends on the behavior of the stochastic field. 

Every stochastic field has a "correlation time" $\tau_t$, defined as the timescale over which the random variable $\xi (t)$ is correlated to itself. Lenstra \cite{Lenstra} works out the ratio $G/J$ required for Born's rule to emerge in SUV model for various limits of the correlation time. It turns out that for all nonzero correlated time,  Born's rule is not satisfied. Only the "uncorrelated" stochastic field leads to emergence of Born's rule such that not only the right measurement statistics are recovered after measurement but also these statistics are conserved during measurement. This conservation of statistics is necessary otherwise if one were to perform an instantaneous measurement whilst the state is collapsing and would be able to measure a different statistics, they could send a signal faster than light. So, the only valid solution is the so-called "white noise limit" where $\tau_t \rightarrow 0$, and the corresponding ratio is $G/J = 1$.


\section{Philosophical Foundations}
Debate on scientific realism is the nexus point of almost all of scientific philosophy for it is concerned with the very nature of scientific knowledge. In this section we provide a brief discussion on realism and anti-realism in connection to QM followed by the implications of SUV for state reduction.  

\subsection{Realism and Anti-Realism in Quantum Mechanics}
Scientific realism is the commitment to a literal interpretation of scientific claims about the world. Semantically, according to a scientific realist, any claims about scientific objects, events, processes, properties, and relations (observable and unobservable), should be construed literally as having truth values, whether true or false. Metaphysically, realism is the belief in the mind-independent existence of the world as described by science.\par
On the other hand, anti-realism refutes the realist's position on one or more of the said claims/commitments: the epistemological commitment to scientific theories to believe some significant portion of what the theory says; the semantic claim of a literal construal of the theories; and the metaphysical commitment to the existence of a mind-independent reality.

\subsubsection{Ontological implications of State Reduction}
\begin{itemize}
    \item SUV attempts to explain collapse in terms of dynamical SSB. Ontologically, this suggests the existence of multiple "possible" symmetry-broken ground states among which the system is "forced to choose" one during state reduction, as seen by the existence of localisation center $x_0$ which depends on the non-unitary, symmetry-breaking perturbation $\mathcal{G}$. This is different from the standard notion of the system somehow "choosing" a ground state on its own from a number of degenerate states. This shift from the system to the perturbation as the deciding factor for symmetry-broken ground state indicates that it might be the stochastic field or the interplay between stochastic field and the state before measurement that decides the post-measurement state. This can be reasoned from the existence of nonlinear term (that depends on the state of the system, which is being measured) and the stochastic term in the non-unitary perturbation. However, one must not rush to the conclusion that since the relative coupling strength is $G/J = 1$ so the system and the stochastic field has an equal influence. We need to remind ourselves that the ratio $G/J = 1$ is valid at uncorrelated limit of SUV and is explicitly defined for two-state model with the non-unitary perturbation given by Eq. \eqref{pert}. 
    \item Perhaps the biggest challenge to traditional ontology of QM is the apparent spontaneous violation of unitarity in large system (thermodynamic limit). The state transitions to symmetry-broken configurations (localized states), suggesting that these are not just mathematical abstractions but reflect physical realities.
\end{itemize}

\subsubsection{Epistemological implications of State Reduction}
\begin{itemize}
    \item The unpredictability inherent in spontaneous symmetry breaking (like localization to $x_0$ in Dynamic SSB) implies limits to our knowledge about the pre-measurement state. Even infinitesimally small perturbations in large systems suffice to "choose" a symmetry-broken ground state, emphasizing the epistemic indeterminacy of quantum states. While true SSB occurs only in the thermodynamic limit, the epistemic treatment of finite systems involves approximate symmetry breaking. This raises questions about the extent to which finite approximations of quantum mechanics capture the true nature of quantum systems.
    \item The SUV framework provides a mechanism for the emergence of Born's rule, linking it to the interplay of stochastic and nonlinear dynamics. This challenges the epistemological stance of quantum mechanics as inherently probabilistic and suggests that measurement statistics emerge dynamically rather than being fundamental.
    \item While true SSB occurs only in the thermodynamic limit, the epistemic treatment of finite systems involves approximate symmetry breaking. This raises questions about the extent to which finite approximations of quantum mechanics capture the true nature of quantum systems.
\end{itemize}

\subsubsection{Semantic implications of State Reduction}
\begin{itemize}
    \item The SUV framework resolves the preferred basis problem by linking it to symmetry-breaking perturbations. The classical-like bases (e.g., position or spin states) naturally emerge as symmetry-broken ground states, avoiding the need for artificial assumptions about preferred bases.
    \item The white noise limit (where $\tau_t \rightarrow 0$) as a necessary condition for Born's rule suggests that QM's probabilistic interpretation is a semantic construct tied to specific dynamic conditions rather than a universal truth. We discuss this more in the next section when we expand on the nature of quantum mechanical probabilities.
    \item The use of non-Hermitian operators and stochastic fields introduces new semantic considerations. For instance, expectation values are redefined to account for time-dependent norms, altering the interpretation of measurement outcomes and dynamics.
\end{itemize}


\section{Nature of Quantum Mechanical Probabilities}
Probability has a mathematical and a philosophical aspect. While the mathematical dimension to probability theory is well-established with mathematicians having almost complete consensus, there is a wide divergence of opinions about the philosophy. And this debate seeps into physics as probability is closely linked to statistics, which is concisely defined by Bas van Fraassen (BvF) as "the science that deals with distributions and proportions (large but finite) in actual classes (also called `populations', `aggregate', `ensembles') of actual things" \cite{BvF}.
A statement about a distribution like \\

\centering About 27\% of physicists worldwide specialize in condensed matter physics.\newline

\justifying can be called a statistic, which is arrived upon by using statistical methods designed to test hypotheses and to infer new statistics from given ones or a data. In doing so, a statistician uses the mathematical theory of probability. According to BvF, philosophical problems arise when introducing probability into physical theories because probability is often interpreted in multiple, potentially conflicting ways, which leads to conceptual confusion. Specifically, the issues stem from the tension between Epistemic vs. Ontic interpretations of probability, or determinism vs. indeterminism theme when trying to reconcile probabilistic descriptions (as in QM) with classical ideas of causation and law-like behavior, which is further elaborated in \cite{BvF}. In this section, we review the nature of probabilities in QM in the light of SUV, and its possible interpretations. 

\subsection{Implications for the Measurement Problem} 
Consider the collapse dynamics for the two-state model based on Lieb-Mattis AFM as described by Eq. \eqref{afm}. As described by Lenstra \cite{Lenstra}, stochastic noise fields provide a way to incorporate randomness and probabilistic effects
into an otherwise deterministic model. In SUV model, the stochastic variable determines
the sign of $\theta (t)$ at a certain time $t$ thereby
determining the flow direction of the state vector on the Bloch sphere, and acts as a repulsive
fixed point. Since the stochastic variable is randomly sampled at different time $t$, it is randomly
determined if the state moves towards the $\ket{0}$ or $\ket{1}$ pole at that time $t$. This is visualized in the flow diagram in figure 3 (taken from \cite{Lenstra}). If $\theta < \eta = \cos^{-1} \xi \text{ with } \theta, \eta \in [0, \pi]$ at a certain time $t$, it follows that $\cos \theta > \xi$, and from Eq. \eqref{afm} we see that then $\Dot{\theta} < 0$. The state
will thus flow towards the $\theta = 0$ point corresponding to the $\ket{0}$ state at that time $t$. On
the other hand if $\theta > \eta = \cos^{-1} \xi$ we get similarly $\cos \theta < \xi \text{ and } \Dot{\theta} > 0$ and we will move
towards $\theta = \pi$, corresponding to the $\ket{1}$ state.

\begin{figure}[htb]
    \centering
        \includegraphics[width=0.4\linewidth]{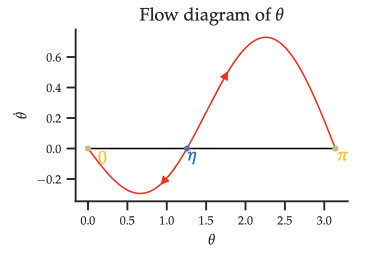}
        \captionsetup{width=\linewidth} 
        \caption{Flow diagram for $\theta \in [0,\pi]$ described by \eqref{afm} with $G = J = 1$ and $\xi(t) = \cos \eta (t)$ with $\eta (t) = 2\pi/5 \forall t$.}
\end{figure}
The arrows here indicate the direction in which $\theta$ will flow, and $\eta$ shows the repulsive fixed point and $0$ and $\pi$ are the attractive fixed points. 
Even though this is explicitly for a two-state system, this proof of principle shows that what determines the post-measurement state is the stochastic variable and the relative coupling strength $G/J$. Whether the state collapses to $\ket{0}$ at fixed point $\theta = 0$ on the Bloch sphere or to $\ket{1}$ at fixed point $\theta = \pi$ depends not just on the value of stochastic variable $\xi(t)$ at the time of collapse but also throughout the measurement process. This changes the typical narration of wave function collapse as now the collapse of wave function itself is a single event occurring at a specific time but the mechanism of collapse that leads the initial state to the final, post-measurement state takes some finite time. This undermines the classical realist view that the measurement problem can be resolved by assuming that quantum states represent pre-existing properties. Instead, the measurement outcomes are seen as emergent phenomenon. It also asserts that if at some time $t' \in [0,t_{collapse}]$ if a second perturbation is turned on then the stochastic the final, post-measurement state is determined by the interplay between the two stochastic fields of the two perturbations. \\ 

In standard quantum mechanics, the measurement problem arises because the wave function evolves deterministically according to the \myS's equation but collapses probabilistically during measurement. SUV introduces a stochastic element explicitly into the dynamics, thereby embedding the probabilistic nature of measurement within the theory itself. The role of the stochastic variable and $G/J$ ratio in the measurement problem suggests a shift from the dualistic nature of the measurement problem (determinism vs. randomness) toward a unified framework where randomness is an inherent, emergent feature of quantum systems. This resonates with ontological indeterminism, implying that nature fundamentally operates in a probabilistic manner, rather than randomness being merely epistemic (a limitation of our knowledge).\par

The ratio $G/J$ determines whether the stochastic term (associated with randomness) or the nonlinear term (associated with deterministic collapse) dominates the dynamics. This context-dependence ties the measurement outcome not just to the system, but also to the broader environmental and experimental setup.

\subsection{Interpreting Quantum Mechanical Probabilities}
Karl Popper has been one of the most influential critics of the Copenhagen interpretation of QM. And he was one of the first modern philosopher of science to take problem of interpretation of probabilities seriously and worked on it. He developed the Propensity interpretation of Probabilities (PIP) to address the problem of interpretation of QM probabilities, especially that of single event, something that frequency interpretation cannot resolve. According to the Frequency interpretation, probabilities are the long-run relative frequencies of outcomes in repeated experiments. They represent empirical regularities observable over many trials.\par
PIP, on the other hand, dictates that probabilities represent objective tendencies or dispositions of physical systems to produce certain outcomes. They are intrinsic properties of the system, not tied to observers or knowledge. PIP has had advocates that claim it to be a conceptually superior alternative to other interpretations like frequentist interpretation, subjectivist interpretation (probabilities represent an agent's degree of belief, and are updated using Bayes' theorem as new information becomes available) or ignorance interpretation (Probabilities arise due to incomplete knowledge or lack of information about the underlying deterministic state of the system. They are epistemic rather than ontological).\\

Shanks \cite{Shanks} argues that single-case PIP, the idea that probabilities represent objective tendencies or propensities of a quantum system to produce specific outcomes in a "single" experiment, rather than as frequencies over multiple trials, is incompatible with relativity. Here we show how SUV resolves the conflict and provides a restatement of PIP.

\subsubsection*{Shanks's argument}
In his paper titled \textit{"Time and the propensity interpretation of probability"}, Shanks provides the following arguments for the incompatibility between PIP and relativity theory:
\begin{itemize}
    \item PIP views QM probabilities as intrinsic tendencies.
    \item This involves a temporal aspect suggesting a direction from potentiality $\rightarrow$ actuality, implying a dynamic view of time.
    \item Relativity, on the other hand, treats time as a dimension similar to space dimensions where the past, present and future are all on an equal footing (no \textit{special} treatment), implying a "block universe".
    \item Conflict between "dynamic" conception of time in PIP with the static, tenseless view of time in relativity.
\end{itemize}

\centering \textbf{Conclusion:} If time does not "flow" and all events are fixed within the spacetime manifold, the idea of propensities actualizing over time becomes problematic.\newline

\justifying Shanks concludes that the disparity between the temporal assumptions of the propensity interpretation and the relativistic model of time presents a significant philosophical challenge. This necessitates a careful reconsideration of how probabilities are conceptualized in quantum mechanics, particularly in contexts where relativistic considerations are essential.

\subsubsection*{SUV to the rescue}
SUV introduces mechanisms and interpretations that could potentially address the incompatibility discussed above:
\begin{itemize}
    \item Dynamic but relational time evolution: SUV accommodates probabilistic outcomes as emergent phenomena that do not require a preferred temporal direction. The collapse dynamics in SUV are contextually tied to the interaction between the system and environment, allowing for a time-neutral explanation that is consistent with relativity’s spacetime framework. In relativity, time is part of a four-dimensional spacetime manifold, and there is no universal "now" or flow of time. SUV bypasses the need for a universal temporal flow by embedding the evolution of quantum probabilities into the system's interaction context, rather than an absolute temporal framework.
    \item Collapse as a local phenomenon: In SUV, the stochastic variable governs the collapse dynamics and depends on localized interactions rather than a global time evolution. This aligns with relativity’s principle of locality, ensuring that probabilistic outcomes are tied to specific regions of spacetime. The probabilities in SUV emerge from the interplay of nonlinear and stochastic terms and do not rely on a dynamic temporal becoming. Instead, they are understood as context-dependent tendencies that reflect the system's configuration and external perturbations.
    \item Avoiding temporal flow: Symmetry-breaking events, whether spatial (as in crystals) or temporal (as in time-translation symmetry breaking), can occur without requiring a universal temporal direction. These events depend on singular limits (e.g., thermodynamic limits or perturbations) that do not privilege any particular temporal framework, making them compatible with relativity. In SUV, probabilities emerge naturally as part of the symmetry-breaking process, which does not require the explicit temporal potential-to-actuality transition assumed by the propensity interpretation. This avoids Shanks’s critique of temporal incompatibility.
\end{itemize}
\centering \textbf{Conclusion:} Probabilities in SUV model of QM reflect intrinsic tendencies (propensities) of quantum systems to evolve toward specific outcomes. However, these tendencies are tied to the stochastic and nonlinear terms in the system’s equations, rather than a temporal dynamic of potentiality to actuality. \newline

\justifying Since SUV models rely on local interactions and do not require a preferred temporal framework, the propensities they describe are fully compatible with relativity. Probabilities emerge as relational properties of systems and their environments, not as absolute temporal processes. SUV resolves the incompatibility by addressing the core issues raised by Shanks: 
\begin{enumerate}
    \item Time Translation Symmetry: Non-unitary dynamics and symmetry breaking in SUV do not assume a preferred direction of time, aligning with the block universe of relativity.
    \item Local Collapse: Stochastic dynamics in SUV respect the locality principle, avoiding faster-than-light (FTL) influences or the need for a global temporal framework.
    \item Emergent Probabilities: SUV derives probabilities as emergent features of the system-environment interaction, consistent with both quantum formalism and relativistic spacetime.
\end{enumerate}
Thus, SUV strengthens PIP arguments by resolving the incompatibility with theory of relativity and provides a restatement of PIP. SUV replaces the temporal potential-to-actuality framework of PIP with a relational, context-dependent model of probabilities. By embedding stochasticity and symmetry breaking into the quantum dynamics, SUV makes probabilistic outcomes compatible with relativity’s block universe and spacetime locality, effectively addressing the incompatibility highlighted by Shanks. This makes SUV a promising approach for integrating quantum mechanics and relativity while preserving an intrinsic understanding of quantum probabilities.  

\subsection{Interpreting Statistical Probabilities}
In order to say something about probabilities in statistical mechanics (SM) one first needs to understand the limiting relation between quantum and classical mechanics. What we get from QM, in an appropriate limiting regime, is the probability distribution (PD) on classical phase space. This distribution then `evolves' approximately like a classical probability distribution. Therefore, the classical limit of QM state is a classical PD.

\subsubsection*{The Wigner Function}
When discussing the correlations between QM and classical statistical mechanics, to link the wave function to a PD in phase space, one encounters the "Wigner function", also called the "Wigner quasi-probability distribution". It maps the functions in the quantum phase space formulation and Hilbert space operators in the \text{\myS } \text{picture}. A quantum state gives us a PD for each set of mutually compatible observables. This allows one to define real valued functions $W(\textbf{x}_1,...,\textbf{x}_n;\textbf{p}_1,...,\textbf{p}_n)$ (for any n-body quantum system) over classical phase space. This so-called "Wigner function" returns a density function for that state's PD over position (momenta), when integrated over all momenta (position). Mathematically, one can write the Wigner function for a mixed state with density operator $\hat{\rho}$ as

\begin{equation}
    W(\textbf{x},\textbf{p}) = \frac{1}{h^3}\int e^{-i\textbf{p}\cdot y / \hbar} \bra{\textbf{x}+\textbf{y}/2}\hat{\rho}\ket{\textbf{x}-\textbf{y}/2}d^3\textbf{y}
\end{equation}
This function now evolves in time corresponding to the evolution of the quantum state. Recall that the classical density function satisfies the \textit{Liouville equation}, 

\begin{equation}
    \frac{\partial \rho_{cl}}{\partial t} + \sum_{i=1}^{N} \left( \frac{\partial \rho_{cl}}{\partial q_i}\frac{\partial H}{\partial p_i} - \frac{\partial \rho_{cl}}{\partial p_i}\frac{\partial H}{\partial q_i}\right) = 0
\end{equation}
where the second term can be written in short as $\{\rho, H\}$ and is called the Poisson bracket. An equivalent expression for $\hat{\rho}$ is given by the \textit{von Neumann equation}:
\begin{equation}
    \frac{\partial\hat{\rho}}{\partial t} = -\frac{i}{\hbar}[\hat{H}_{SUV},\hat{\rho}]
\end{equation}
Using Eq. (20) and Eq. (22) one can write the evolution of Wigner function as:
\begin{equation}
    \frac{\partial W}{\partial t} = \{H_{SUV},W\} + \mathcal{Q}(W)
\end{equation}
where, $H_{SUV}$ is the Weyl transform of $\hat{H}_{SUV}$ which is simply the Hamiltonian function with operators replaced by corresponding variables (this is true because the terms in $\hat{H}_{SUV}$ do not involve any cross terms in $\hat{P}_{tot}$ and $\hat{X}_{com}$ \footnote{Check \cite{wigner} for an introduction to Weyl transforms and Wigner function}), and $\mathcal{Q}(W)$ includes quantum corrections given by:

\begin{equation}
    \mathcal{Q}(W) = \sum_{n=1}^{\infty} \frac{ (-\hbar/2)^{2n}}{(2n+1)!}\frac{\partial^{2n+1}U}{\partial q^{2n+1}}\frac{\partial^{2n+1}W}{\partial p^{2n+1}}
\end{equation}\\
where $U$ is the potential term in the Hamiltonian (in our case it simply has a coefficient $i$, or equivalently, we replace $U$ with $\mathcal{U} \equiv i\epsilon\mathcal{N}\mathcal{G}$. The exact, rigorous derivation of Eq. (24) in this case doesn't matter because we show that $\mathcal{Q}(W)$ vanishes ultimately).
Computing the evolution of Wigner function for the Hamiltonian in Eq. \eqref{Hsuv} we find that
\begin{equation}
    \frac{\partial^{2n+1}\mathcal{U}}{\partial X_{com}^{2n+1}} = \begin{cases}
        0, & \text{ if $2n+1 > 2$}\\
        2i\epsilon\mathcal{N}, & \text{ if $2n+1 = 2$}
    \end{cases}
\end{equation}
Since the potential is quadratic in $X_{com}$, and $n \in \mathbb{N}$,  the above expression implies $\mathcal{Q}(W) = 0$. Therefore, one can simply compute the Poisson bracket in Eq. (23) and get:
\begin{equation}
    \frac{\partial W}{\partial t} = 2i\epsilon \mathcal{N} (X_{com} - x_0)\frac{\partial W}{\partial P_{tot}} - \frac{P_{tot}}{m_{tot}}\frac{\partial W}{\partial X_{com}}
\end{equation}
Taking the classical limit of this ($\hbar \rightarrow 0$) reduces the Wigner function to the classical probability density. However, before taking that limit we see the effect of taking the two limits of Eq. \eqref{l1} and Eq. \eqref{l2}:

\begin{equation}
    \lim_{\mathcal{N}\rightarrow \infty}\lim_{\epsilon \rightarrow 0} W(X_{com}, P_{tot}) = W_{loc}(X_{com}, P_{tot} = 0) 
\end{equation}
where $W_{loc}(X_{com}, P_{tot} = 0)$ is Wigner function localised in momentum space at $P_{tot} = 0$. One can see, from Eq. \eqref{l1}, that in the above limit $P_{tot} \rightarrow 0$ and the second term vanishes. As $\epsilon \rightarrow 0$, the first term also vanishes indicating $\partial W/ \partial t = 0$, thus a stationary Wigner function, consistent with localisation in momentum space.\par
Similarly, for the limit of Eq. \eqref{l2}, we find:
\begin{equation}
    \lim_{\epsilon \rightarrow 0}\lim_{\mathcal{N}\rightarrow \infty} W(X_{com},P_{tot}) = W_{loc}(X_{com} = x_0, P_{tot}) 
\end{equation}
where $W_{loc}(X_{com} = x_0, P_{tot})$ is the Wigner function localised in position space at $X_{com} = x_0$. Again, from Eq. \eqref{l2}, one can see that in this limit $X_{com} \rightarrow x_0$, vanishing the first term. As the localisation takes place in the position space, $P_{tot} \rightarrow 0$; thereby vanishing the second term. This leads to stationary condition $\partial W/ \partial t = 0$.\par
In the classical limit, the evolution equation of Wigner function would lead to the following Liouville equation for the classical probability density function:
\begin{equation}
     \frac{\partial \rho_{cl}}{\partial t} = 2i\epsilon \mathcal{N} (X_{com} - x_0)\frac{\partial \rho_{cl}}{\partial P_{tot}} - \frac{P_{tot}}{m_{tot}}\frac{\partial \rho_{cl}}{\partial X_{com}} 
\end{equation}
The first term here arises from the symmetry-breaking perturbation, introducing an effective "diffusive-like force" in momentum space, whereas, the second term is the classical transport term describing the free evolution of the system in phase space. 

\subsubsection*{Reflections}
The Wigner function shows how QM and SM share a unified probabilistic framework, with SM probabilities naturally arising from the classical limit of quantum processes. Thus, it serves as a bridge between QM and SM, showing that the probabilistic structures of SM are a natural continuation of quantum stochastic dynamics in the classical limit. But one can't just suppose it to be a classical PD; as Wallace \cite{wallace} points out that \textit{"there is no underlying microdynamics on phase-space points such that the Wigner function's dynamics can be seen as a probabilistic dynamics for that microdynamics"}.\\

However, it not entirely useless either; in regimes where the higher-order terms in its evolution can be neglected (like the example above), and in which the system is not probed on length-scales where the uncertainty principle comes into play (valid in the mesoscopic regime in the above example), there is an approximate isomorphism between the dynamics of the quantum state and that of the classical PD \cite{wallace}. In such cases, SUV aligns with Wallace’s interpretation that SM probabilities are rooted in QM probabilities. It provides additional insight into the quantum-classical transition in phase space by introducing stochastic and symmetry-breaking dynamics that reinforce the emergence of classical probabilities and temporal asymmetry. Here are some insights provided by SUV:

\begin{enumerate}
    \item Nonunitarity in Phase Space: SUV introduces non-unitary dynamics (via Lindblad-like terms) into the density matrix. These dynamics, when mapped to phase space, appear as stochastic and dissipative effects in the Wigner function evolution. This provides a mechanism for how interference is suppressed and classicality emerges.
    \item Symmetry Breaking and Localization: SUV emphasizes the role of symmetry breaking in selecting definite outcomes (e.g., a pointer state). In phase space, this corresponds to the Wigner function localizing into a classical probability distribution.
    \item Temporal Asymmetry: In SM, non-equilibrium systems evolve toward equilibrium, with probabilities tied to irreversible processes. SM probabilities reflect irreversible, non-equilibrium dynamics with a directional flow of time analogous to how the dynamic view of time from potentiality $\rightarrow$ actuality in SUV-based PIP describes the evolution of a quantum system from a delocalised, symmetric quantum state (e.g., superpositions in phase space or Hilbert space) to a localised, symmetry-broken state (e.g., a definite position, momentum, or pointer state). This process is irreversible due to stochastic and dissipative effects, analogous to the arrow of time in SM.
\end{enumerate}
The last point above is perhaps the \textit{key insight} provided by SUV on the interpretation of the nature of SM probabilities: irreversible processes transform the system from a symmetric, non-equilibrium configuration to a symmetry-broken, stable one:
\begin{itemize}
    \item SUV: Non-unitary dynamics evolve the system from a symmetric quantum superposition to a classical-like definite state.
    \item SM: Temporal asymmetry drives the system from non-equilibrium (low entropy) to equilibrium (high entropy), breaking time-reversal symmetry.
\end{itemize}
Both frameworks explain emergent, probabilistic behavior as a consequence of underlying irreversible dynamics. And SUV provides a physical mechanism for how SM probabilities inherit their quantum origins, reinforcing Wallace's interpretation.\par
Wallace's and the SUV framework's positions align with an ontic perspective, proposing that the probabilities in SM are inherently quantum mechanical. But this doesn't mean that classical SM cannot stand on its "own two feet". As mentioned by Myrvold \cite{myrvold}, introducing an unspecified source of uncertainty is sufficient for SM to function effectively, implying that the foundational basis of this uncertainty is not crucial for the application of SM probabilities. Myrvold's stance can be seen as adopting an epistemic or pragmatic approach, where the source of uncertainty in SM is treated as a practical tool for prediction and explanation, without necessitating a quantum foundation. The divergence between Myrvold and Wallace (and by extension, the SUV framework) is rooted in differing philosophical interpretations of the origin and nature of probabilities in SM.\\

In summary, the debate between Wallace's ontic interpretation and Myrvold's epistemic stance reflects the broader philosophical tension between reductionism and independence in the interpretation of SM probabilities. While Wallace and the SUV framework argue for the quantum mechanical origins of SM probabilities, emphasizing the role of stochasticity, symmetry breaking, and temporal asymmetry, Myrvold highlights the pragmatic sufficiency of SM probabilities without requiring quantum underpinnings. Both perspectives offer valuable insights: Wallace’s approach aligns with a unified probabilistic framework bridging QM and SM, while Myrvold’s stance affirms the utility and autonomy of classical SM as a predictive and explanatory tool. Ultimately, the choice between these interpretations depends on whether one seeks a foundationally unified framework or prioritizes practical independence in describing statistical mechanics.

\subsection{Critiques and Limitations of SUV models}
Finally, it is worth mentioning some limitations and methodological concerns regarding SUV models:
\begin{itemize}
    \item The uncorrelated SUV model involves white-noise in state space $\mathcal{H}$ which is difficult to justify, since this would imply that for any $\mathcal{N}$, the underlying process resulting in the non-unitary evolution would have a negligible characteristic time. So, whatever process causes collapse would have to vary over a timescale smaller than all timescales known so far in physics. This is not impossible but rather implausible.
    \item While constructing SUV, one could also have looked at the stochastic field which is not white noise and yet reproduces Born's rule and shows correct large system localisation behavior. However, this is yet no explored as it requires the mathematical theory of noise homogenization which is quite restricted to real stochastic process. 
    \item A rather apparent, fundamental problem is the origin of the stochastic field. There is no justification or proof for the nonunitary perturbation being the only perturbation resulting in Born's rule and localisation behavior. 
    \item Just as \myS's equation does not describe the behavior of superpositions of states of massless particles, so does SUV models. The time axis over which the evolution takes place is ambiguous in this case due to absence of a proper time.
    \item At last, it is important to mention that for this model to agree with causality one needs Born's rule to hold during measurement. To implement this either the stochastic field has to exist in white noise limit which leads to the issues mentioned in the first point above, or one has to give up the assumption that FTL signaling during collapse immediately leads to causality violations. In other words, one might have to redefine collapse in a global, instead of a local reference frame (owing to the non-local nature of QM) or otherwise, critically analyze whether it is influences or correlations that travel through the closed space-like loop thus created.
\end{itemize}



\section{Conclusion}

The measurement problem illustrates our incomplete knowledge of the mechanism of wave function collapse. We do not know precisely when it occurs and what causes it; and the boundary between quantum and classical physics remains blurry. However, with SUV models we get some intuition about a way forward. The most significant one of them being that macroscopic wave function collapse is a result of SSB of the unitary time evolution of standard QM's \myS's equation, and it is the perturbation (or rather the interplay between the quantum state and the stochastic field) that "forces" the system to choose a symmetry-broken state. Here we provide an outline of this mechanism and discuss its ontological, epistemological and semantic implications. Note that Born's rule emerges in SUV models and is not added as an axiom or is not a result of fine-tuning of parameters. Also, SUV solves the preferred basis problem by showing that classical-like bases naturally emerge as symmetry-broken ground states. 

Next, we discussed the nature of probabilities in QM and their interpretation. We notice how the collapse mechanism as described by SUV aligns with the realist interpretation as the collapse process is brought out by the stochastic field(s) (and its interplay with the state being measured; embedded in the nonlinear term), in some small, non-vanishing timescale which can be tested given a sensitive measurement machine \cite{test}. 
SUV allowed us to resolve the conflict between PIP and the theory of relativity thereby strengthening the PIP argument and, at the same time, providing it a stochastic aspect. This modified PIP argument supports the emergent view that probabilities arise in the thermodynamic limit as a result of stochastic dynamics and symmetry-breaking. This interpretation was then linked to SM probabilities via the use of Wigner function, where we explicitly showed its implications for a crystal with harmonic, symmetry-breaking perturbation. We find that SM probabilities have its roots in QM probabilities and SUV provides a unified framework for this. Based on these conclusions one can draw a relation between QM and SM: \par

\begin{center}
\vspace{0.3in}
\begin{tikzpicture}
    \draw[->] (-4,0) -- (4,0);
    \draw[->] (-4,-4) -- (4,-4);
    \draw[dashed] (-7,-0.5) -- (-7,-3.5);
    \draw[dashed] (-5.5,-1) -- (-5.5,-3);
    \draw[dashed] (0,-1) -- (0,-3);
    \draw[dashed] (5.8,-1) -- (5.8,-3);

    \node at (-7,0) {QM:}; 
    \node at (-5.5,0.4) {Delocalised,};
    \node at (-5.5,0) {symmetric};
    \node at (-5.5,-0.4) {state};

    \node at (5.8,0.4) {Localised,};
    \node at (5.8,0) {symmetry-broken};
    \node at (5.8,-0.4) {state};

    \node at (-1.2,1.1) {Breaking};
    \node at (-1.2,0.7) {time-translation};
    \node at (-1.2,0.3) {symmetry};
    \node at (0.3,0.7) {\huge /};
    \node at (1.6, 0.7) {Non-unitarity};
    \node at (-0.1, -0.4) {mechanism: (Non-Eq.) DSSB};

    \node at (-7,-4) {SM:}; 
    \node at (-5.5,-3.8) {Non-Eq.};
    \node at (-5.5,-4.2) {state};

    \node at (5.8,-3.8) {Equilibrium};
    \node at (5.8,-4.2) {state};

    \node at (-1.5,-3.7) {Temporal asymmetry};
    \node at (0.4,-3.6) {\Large /};
    \node at (1.7, -3.7) {Irreversibility};
    \node at (-0.1,-4.3) {mechanism: Non-Eq. dynamics};
    \node at (0,-6.5) {Figure 4: Relation between QM and SM};
    
\end{tikzpicture}
\end{center}

\end{document}